\documentclass[prl,twocolumn,floatfix,epsfigs,nofootinbib]{revtex4}
\usepackage{amssymb,amsmath,amsfonts}
\usepackage{graphics,dcolumn,bm,fleqn,epic,eepic,float}
\usepackage{graphicx} % Include figure files
\usepackage{epstopdf} % auto-convert eps to pdf
\usepackage{bm} % bold math
\usepackage{subfigure}
\usepackage{color} %  for colored comments
\begin{document}
\title{Boosting test-efficiency by pooled testing strategies for SARS-CoV-2}

%\author{Rudolf Hanel$^{1,2}$, Mathias Beiglb\"ock$^{3}$, Stefan Thurner$^{1,2,4,5}$}
\author{Rudolf Hanel$^{1,2}$, Stefan Thurner$^{1,2,3,4}$}

\affiliation{%
  $^1$~Section for Science of Complex Systems, Medical University of
  Vienna, Spitalgasse 23, 1090 Vienna, Austria}
   \affiliation{%
  $^2$~Complexity Science Hub Vienna, Josefst\"adterstrasse 39, 1080 Vienna, Austria}
 %\affiliation{%
 % $^3$~Faculty of Mathematics, University of Vienna, 1090 Vienna, Austria}
 \affiliation{%
  $^3$~Santa Fe Institute, Santa Fe, NM 87501, USA}
\affiliation{%
  $^4$~IIASA, Schlossplatz 1, 2361 Laxenburg, Austria}
\date{\today}

\begin{abstract} 
In the current COVID19 crisis many national healthcare systems are confronted with an acute shortage of tests for confirming SARS-CoV-2 infections. 
For low overall infection levels in the population, pooling of samples can drastically amplify the testing efficiency. 
Here we present a formula to estimate the optimal pooling size, 
the efficiency gain (tested persons per test), and 
the expected upper bound of missed infections in the pooled testing,
all as a function of the population-wide infection levels and the false negative/positive rates of the currently used PCR tests. 
Assuming  an infection level of $0.1\ \%$  and a false negative rate of $2\ \%$, the optimal pool size is about $32$, 
the efficiency gain is about $15$ tested persons per test. 
For an infection level of  $1\ \%$ the optimal pool size is $11$, the efficiency gain is $5.1$ tested persons per test. 
For an infection level of  $10\ \%$ the optimal pool size reduces to about 4, the efficiency gain is about $1.7$ tested persons per test. 
For infection levels of  $30\ \%$ and higher there is no more benefit from pooling. 
To see to what extent replicates of the pooled tests improve the estimate of the maximal number of 
missed infections, we present all results for 1, 3, and 5 replicates.
%Finally, we demonstrate that consecutively pooling a positively tested group does not increase the efficiency gain.
\end{abstract}

\date{March 21, 2020}
%\keywords{quantitative social science | gender | multiplex networks | network theory | online games}

\maketitle

\section{Introduction}
We briefly analyse how \textit{pooled testing} increases the efficiency %and coverage 
of virus testing, given that only a limited number of tests is available. 
The idea is to pool samples taken from several subjects 
and test the combined sample with a single test. 
If the test is negative all subjects are negative. 
If the test is positive all individuals are tested to find the infected ones.
%Alternatively, the positively tested group of subjects could be subdivided into sub-groups, which are then subsequently tested, each with one test.  
The idea was recently suggested by Dina Berenbaum of the Technion Israel Institute of Technology \cite{israel1} and 
has been implemented at the Rambam Medical Center and the Technion in Haifa, who invite other hospitals to follow their example  \cite{israel2}.
Initial tests there indicate that pooling works for the SARS-CoV-2.
It was suggested that up to 32---maybe even 64---people could be tested with a single test, 
which could help to significantly reduce the efforts of mass testing at hospitals and airports.

Here we contribute an estimation of the benefits of such  pooling strategies and the optimal pooling size. 
The optimal size depends on the fraction of infected in the population and the rates of 
false positives and false negatives of the tests in use. 
We provide a formula for the optimal pooling size, the effective number of tested persons per test, and the 
false negatives for the pooled test. 
Our conclusion is that given current infection levels (as of March 21, 2020) 
the pooling size should be smaller than the suggested 32-64.

\section{Model}

We assume that 
\begin{itemize}
\item 
a fraction $\lambda$ of infected people in the population, 
\item
tests have a \textit{false positive} rate of $\gamma_+$ and a \textit{false negative} rate  of $\gamma_-$. 
We assume that testing a pooled sample does not change the false positive and false negative rates of the test, 
\item
we pool samples into groups of size $\omega$,
\item
to control false negatives we take $r$ replicates of the pooled test.
We then apply a majority rule, meaning that if the majority of the $r$ replicates are positive, the pooled sample is declared positive, 
\item if the pooled test is positive we test each individual in the group separately.  
%Strategy (2) {\em consecutive pooling}: the group is sub-divided into two groups, which then are separately tested. In any of those sub groups that tests positive all individuals get tested individually. In our model we assume that at least one of the subgroup tests positive, while the second subgroup tests positive with the probability of a group half the size of the original group.
\end{itemize}

Under these assumptions we compute  

\begin{itemize}
\item the optimal group size, $\omega^{\rm opt}$
\item  the effective number of persons that can be tested with one test, $PPT$ (persons per test), and  
\item an estimate for the upper bound for the fraction of infected individuals 
that are missed by the pooled testing procedure (applied to the population). 
We call it  the {\em false negative factor for pooled testing} and denote it by $FNPT$.   
We also compute the false negative rate, $FNR$, of pooled testing, which is the fraction of infected 
individuals the pooled test will miss. 
\end{itemize}

\begin{figure}[t]
	\centering
	 \includegraphics[width=0.49\columnwidth]{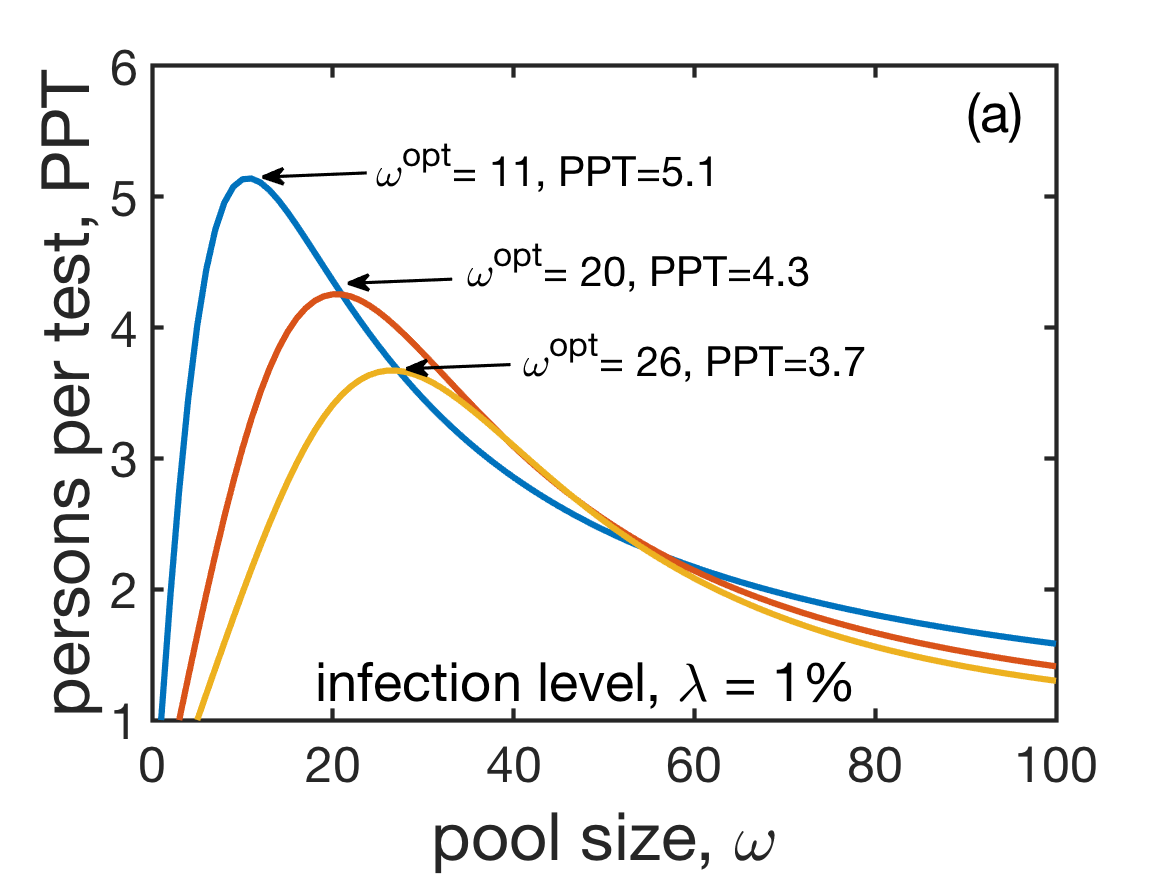}
	 \includegraphics[width=0.49\columnwidth]{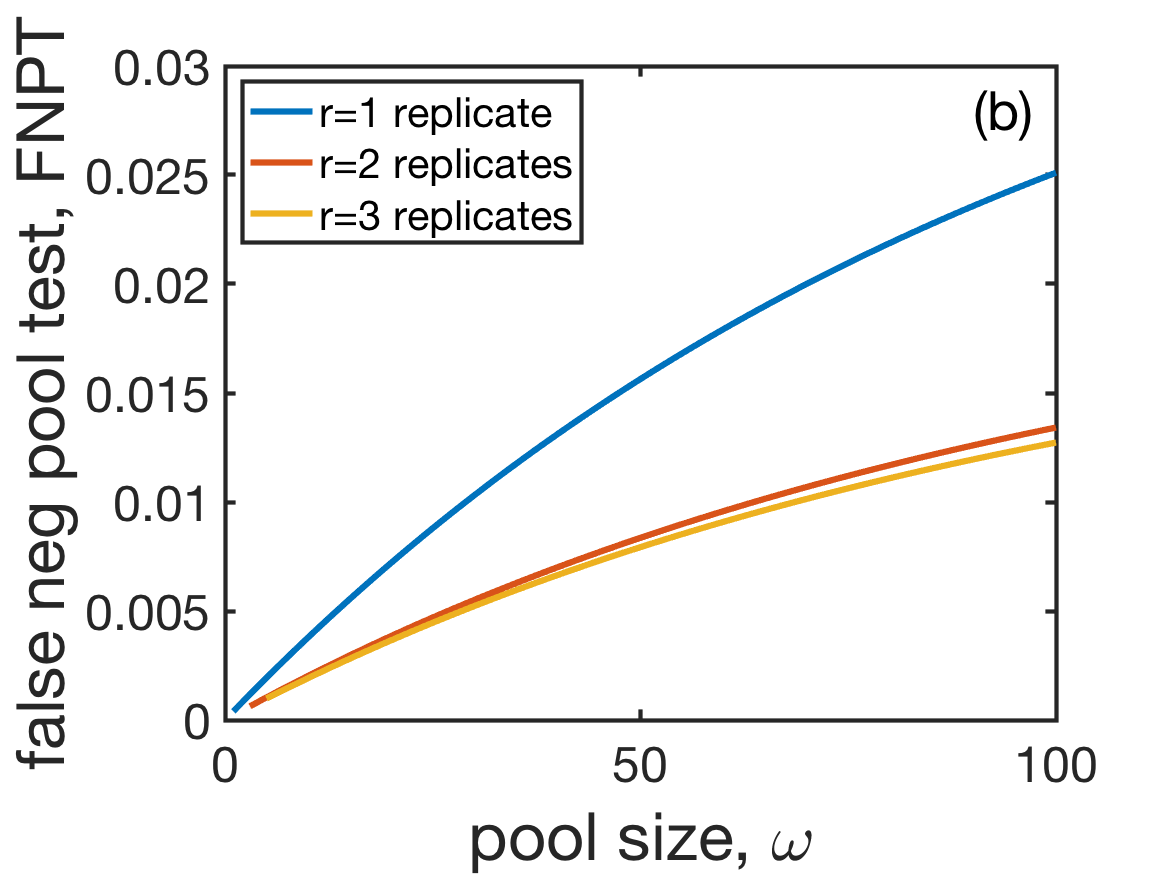}
	\caption{(a) Increase of test efficiency in persons per test, $PPT$. 
		The maximum of this curve indicates the optimal pool size, $\omega^{\rm opt}$
		for a given infection rate, and given false negative and positive rates of the test. 
		Results are shown for $r=1,3,$ and $5$ replicates of testing the pooled sample. 	  	
		the maximum efficiency gain is naturally found for $r=1$ and is about $5.1$ persons per test. 
		(b) False negative factor for the pooled sample, $FNPT$, in \%. 
		The result shows that taking more replicates decreases the false
		negatives. However, note that this also decreases the efficiency to $4.3$ $PPT$.   
		 $\gamma_+=0.0012$ and $\gamma_-=0.02$.
	}
	\label{fig:poolsizes}
\end{figure}

We call a group \textit{positive} if at least one of its members is positive. 
The probability of a group to be positive is 
\begin{equation}
p = 1-(1-\lambda)^\omega \, .
\end{equation}
Because of false positives and false negatives, tests will be positive in $(1-p)\gamma_+$ cases. 
False positives do not decrease the chances to capture a true positive but only decrease the efficiency in using the available tests. 
More importantly, tests will miss positive individuals in $p\gamma_-$ cases on average.   
Hence,  the probability, $P_+$, that a test shows {\em positive}  is 
\begin{equation}
P_+=p(1-\gamma_-)+ (1-p)\gamma_+ = p(1-\gamma_+-\gamma_-)+\gamma_+\, .
\end{equation}
\begin{figure*}[htb]
	 \includegraphics[width=2.2\columnwidth]{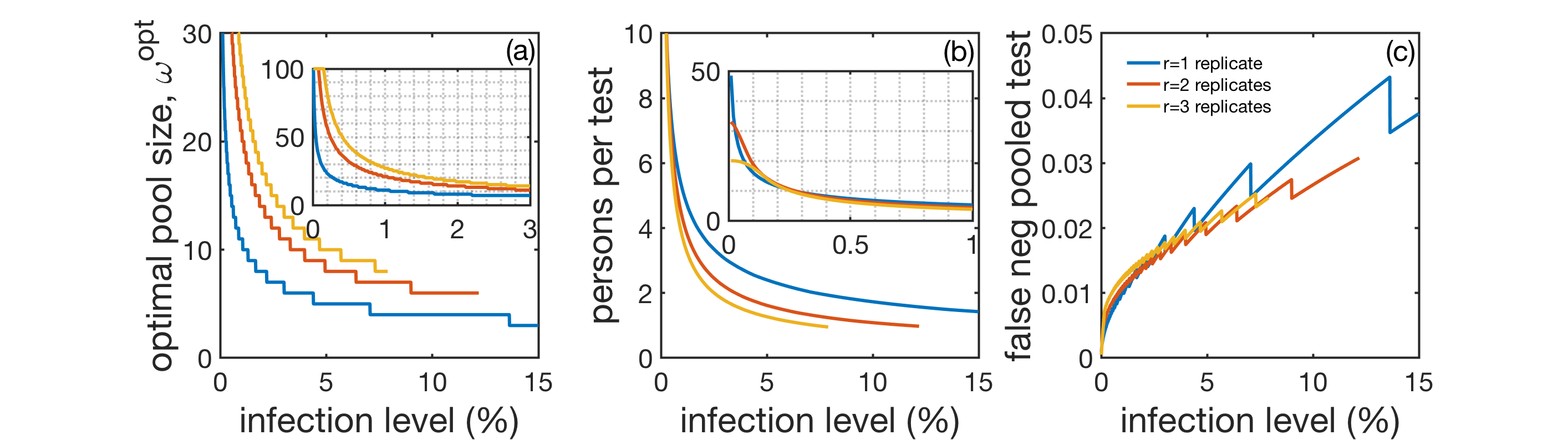}\\
	\caption{(a) Optimal pool size, $\omega^{\rm opt}$, as a function of the infection level of the population.
	The inset is a blowup for low infection levels. 
	The cases for $r=1,3,$ and $5$ replicates is shown in blue, red, and orange, respectively. 
	(b) Efficiency gain of persons per test, $PPT$; the inset shows low infection levels.
	(c) False negative factor for the pooled sample, $FNPT$, in \%. 
	It is clear that taking more replicates does practically not lower $FNPT$. 
%	Note that the zigzag of the curve comes from discrete jumps of the pooling sizes.
	$\gamma_+=0.0012$ and $\gamma_-=0.02$. 
	By taking  $\gamma_-=0.05$, $\omega^{\rm opt}$ and $PPT$ remain practically unchanged, 
	$FNPT$ doubles for all infection levels in this case (not shown).
	}
	\label{fig:2}
\end{figure*}
To see how replicates effect the false negatives of the pooled test we take $r$ replications   
and then apply the majority rule. The probability $P^*_+$ that the majority of $r$ replicates identifies the group as positive becomes 
\begin{equation}
P^*_+=\sum\limits_{i>r/2} {r \choose{i}} P_+^i(1-P_+)^{r-i}\, .
\end{equation}
The expected number of tests per person, $Q$, is therefore  
\begin{equation}
Q=\frac{1}{\omega}\left(r(1-P^*_+)+(r+\omega)P^*_+\right)=P^*_++\frac{r}{\omega}\, ,
\end{equation}
and the {\em persons per test} is simply, $PPT=1/Q$.
Similarly, we can compute an upper bound for the expected number of cases that we miss in pooled testing, $FNPT$. 
It is expressed as the expected number of missed infections per tested person (not per tested infected person). 
If we assume that a group is positive and we test it, then we miss it when the majority rule gives us a negative, 
which happens, when in the majority of cases we get a false negative. 
We therefore get that 
\begin{equation}
	FNPT \equiv p (\gamma_-^*+(1-\gamma_-^*)\gamma_- )\ .
\end{equation}
with
\begin{equation}
	\gamma_-^*=\sum\limits_{i \geq r/2} {r \choose{i}} \gamma_-^i(1-\gamma_-)^{r-i} \ .
\end{equation}
Note that the expected {\em maximal} number of missed infections, $FNPT$, must 
not be confused with the {\em expected false negative rate}, 
$FNR=\gamma_-^*+(1-\gamma_-^*)\gamma_- \sim 2\gamma_-$,  
which is the {\em average} number of individuals one expects to miss in 
pooled testing on average per infected person\footnote{Note that the approximation holds for $r=1$ only and small $\gamma_-$.}. 
If there are no biases or correlations within or between groups, we get that the 
number of missed infections will be 
$\lambda \ FNR \sim 2\gamma_-\lambda$, which does not depend on the pooling size. 
The advantage of $FNPT$ over $FNR$ is that in testings of {\em biased groups} one can be 
confronted with correlated cases with an increased chance of multiple infections within a group, w.r.t. the entire population. 
$FNPT$ therefore captures this situation by considering the {\em upper bound} rather than the average. 
It can be easily checked that $FNPT$ is proportional to $\omega \ \lambda \ FNR$. 

%{\bl 
%For the two stage test we proceed by assuming that we test each group only once ($r=1$).
%We assume that if the initial group tests positive, then at least one of the subgroups tests positive too.
%We assume that the second group tests positive with a probability $p_2=p(\omega/2)$.
%With $P_2=p_2(1-\gamma_+-\gamma_-)+\gamma_+$  we obtain
%\begin{equation}
%Q=\left(1+P_+(2+(1+P_2)\omega/2)\right)/\omega\ ,
%\end{equation}
%and again $PPT=1/Q$.
%Similarly we get an estimate for the $FNPT$ that 
%\begin{equation}
%% um diesen lambda factor geht's
%%FNPT= \lambda p \left(\gamma_-+(1-\gamma_-)\gamma_-(1+P_2)/2\right)\ .
%FNPT= p \left(\gamma_-+(1-\gamma_-)\gamma_-(1+P_2)/2\right)\ .
%\end{equation}
%}

\section{Results}

Results for the optimal pooling size, $\omega^{\rm opt}$, and persons per test, $PPT$, 
are shown in Fig. \ref{fig:poolsizes} (a) for a population-wide infection level of 1 \%.
In Fig. \ref{fig:poolsizes} (b) the increase of $FNPT$ with pooling size is seen. 
Here we use  a false negative rate of $\gamma_-=0.02$ and a false positive rate of 
$\gamma_+=0.0012$, which are sensible estimates for PCR tests that are currently used in Austria (as of March 20) \cite{puchhammer}.
We show the case for $r=1, 3,$ and $5$ replicates for the pooled test in blue, red, and orange, respectively. 

In Fig. \ref{fig:2} (a) we see the optimal pool size, $\omega^{\rm opt}$, as a function of the infection level of the population.
The inset shows the case for low infection levels between $0$ and $3 \ \%$. 
The case for $r=1,3,$ and $5$ replicates is shown in blue, red, and orange, respectively. 
Figure \ref{fig:2} (b) shows the optimal efficiency gain of persons per test, $PPT$, with an inset for small infection levels.
Figure \ref{fig:2} (c) gives $FNPT$ for the pooled testing. It is clearly visible 
that taking more than one replicates does practically not improve the situation (lower $FNPT$). 
Note that the zigzag of the curve comes from discrete jumps of the pooling sizes.

We computed the same values for a false negative rate of $\gamma_-=0.05$.
The results for  $\omega^{\rm opt}$ and $PPT$ practically do not change, 
however, in this scenario, the $FNPT$ approximately doubles for all infection levels. 

%{\bl Strategy (2) we compute numerically, and present it in Fig.  \ref{fig:3}  (red), 
%in comparison to strategy (1), where {\em all} subjects of a positive pooled sample are tested.
% Figure  \ref{fig:3}  (a) indicates that the optimal pool size for strategy (2) is about  a factor 1.4 higher than for strategy (1). 
% The efficiency gain is practically unchanged; see Fig.  \ref{fig:3} (b),  
 %the false positives for the pooled sample for strategy (2) are slightly above those for strategy (1), as seen in Fig.  \ref{fig:3} (c). 
% }
%\begin{figure*}[htb]
%	 \includegraphics[width=2.2\columnwidth]{figure3.png}\\
%	\caption{Comparison of strategy (1) where all subjects of a positive pooled sample are tested (blue)
%		and strategy (2), where the positive sample is again pooled into sub-groups (red). 
%	   Same panels as in Fig 2. 
%	   (a) It turns out that the optimal pool size for strategy (2) is about  a factor 1.4 higher than for strategy (1). 
%	   (b) The efficiency gain is practically unchanged.
%	   (c) The false positives for the pooled sample, $FNPT$, for strategy (2) are slightly above those for strategy (1). 
%	   }
%	\label{fig:3}
%\end{figure*}

\section{Conclusions}

\begin{itemize}

\item The optimal pool size and efficiency of pooling strongly depends on the infection level of the population.  
Let's assume the simplest case of only one test (1 replicate).
From Fig. \ref{fig:2} (a) and (b) we read off that for an infection level of $0.1\ \%$, the optimal pool size is about $32$, 
the efficiency gain is about 15 tested persons per test. 
For an infection level of $1\ \%$, the optimal pool size is 11, 
the efficiency gain is about $5$ fold. 
For an infection level of $10\ \%$, the optimal pool size is reduced to 4, 
the efficiency gain is a factor of $1.7$. 
For infection levels of $15\ \%$ this factor drops below $1.5$ and the optimal pooling size becomes 3. 
This size of 3 remains the optimal pooling size up to infection levels of 29 \%  where the efficiency drops to $1.1$.
From infection levels of 30 \% and larger pooled testing ceases to be effective.  

\item Replicates should help to lower the false negative factor for the pooled strategy, $FNPT$.
In fact, this is only warranted for infection levels larger than $2\ \%$. 
Overall, the use of more than one replicate is clearly not indicated. 
We estimate that for one replicate at an infection level of 
$0.1\ \%$ we will miss about $1$ case in $800$ ($0.13\ \%$) at most. 
At $1\ \%$ we will maximally miss about $1$ case in every $241$ pooled tests ($0.41\ \%$); see also Fig. (\ref{fig:poolsizes}) (b).
Finally, $FNPT$ does not depend strongly on the number of replicates, while optimal pool size depends 
strongly on the number of replicates. $FNPT$ increases with the false negative rates of the test, $\gamma_-$. 
%{\bl
%\item[(c)] Consecutive pooling practically does not gain an extra benefit in terms of tested persons per test, regardless of infection levels. 
%It increases the optimal pool sizes by a factor of somewhat below 1.4.
%}
\end{itemize}
Let us emphasize that a pooling strategy is most powerful for population-wide screening, for example at airports. 
To use them for highly biased samples, e.g. for samples from patients showing symptoms, 
will obviously be much less effective.  

\section{Example}
We finish with a practical example. 
For Austria, a country with slightly less than 10 million inhabitants an actual infection level 
of $0.1\ \%$ would indicate an optimal pool size of 32. For a level of $1\ \%$ it would be 11. 
Assuming the true number of infected to be somewhere between 10,000 and 100,000 this would 
mean a reasonable choice of pooling sizes of about 20.
This number is definitively lower than the suggested sizes reported in \cite{israel1,israel2}. 
The expected gain would be about a factor of 10.  
The use of 1 replicate is indicated.
\\

\begin{acknowledgments}
\subsection{ACKNOWLEDGEMENTS}
We thank Mathias Beiglb\"ock, Walter Schachermayer,  and Reinhard Winkler for 
a careful reading of the manuscript, helpful discussions, and suggestions for future improvements.
This work was supported in part by the Austrian Science Promotion Agency, FFG project under  857136. 
\end{acknowledgments}

%{\bf Author contributions.}
%RH and ST conceptionalized the work, interpreted the results, and wrote the paper. 
%RH designed and coded the model. MB suggested the topic, read, and approved the manuscript.
%The authors declare no competing financial interests.
%The authors are part of the Corona Collaboration at the CSH Vienna.  
 
%\section{Additional information}
%\subsection{Competing financial interests}
%The authors declare no competing financial interests.

%\bibliographystyle{naturemag}
%\bibliography{bibszell2011pardussocgender1}

\end{document}